\documentstyle{article}
\begin{document}

{\bf \LARGE What is research?\footnote{Modified version of the text published
originally in Metaphysical Review, 4(2), 5 (1997), an electronic journal which
has disappeared.}}

\vspace{6mm}

{\Large Mart\'\i n L\'opez Corredoira}

Instituto de Astrof\'\i sica de Canarias

C/.V\'\i a L\'actea, s/n

ES-38200 La Laguna (Tenerife), SPAIN

E-mail: martinlc@iac.es

\vspace{15mm}

I am going to take a critical look at what it means to do research. The point 
of view I express is a personal one that seeks to disturb the calm and 
complacent consciences of those who are dedicated to research. You may 
not agree with me, but I hope that at least you will take what I say as 
an invitation to reflect and perhaps even formulate fresher points of view.

It all began when our primary or secondary school teachers inspired us by 
telling stories about the struggles and miracles of science. This did not 
impress all the pupils, but it did some of us-after all, here we are doing 
research. Our ability to solve problems that the other kids found difficult 
was a factor that sharpened our interest in the sciences. We felt a 
kinship with science and mathematics and experienced a certain 
ego-boosting pride, as if to say, ``here I count for something''.

As lovers of scientific knowledge, simply thinking about the great events 
in scientific history was enough to feed our intellects. We were told of 
the exploits of Galileo, Newton, Darwin, Einstein or Bohr, who for us became 
heroes worthy of emulation.

Eventually, we finished our degrees with high grades and were able at last to 
gain access to one of those 'high-tech' centers where, so we were told, 
research is done. ``But what exactly is it that we really do?'' 

History teaches us not to separate individual experience from general events. 
Hence, when Galileo observed the satellites of Jupiter, he was also 
demonstrating that not everything revolves around the Earth. Newton's 
laws were not formulated for the purposes of engineering applications 
but to reveal the non-teleological mechanism of our Universe. When Laplace 
told Napoleon that he had solved the system of equations which explain 
the motion of the planets without the need to invoke God, the important 
point was not his mathematical juggles but the struggle to arrive at the 
truth without resorting to ancient mythologies. Darwin put humankind in 
its place within the animal kingdom. Etc.

Changing ideas about the Universe are what drives the scientist. Before 
becoming a data collection on Nature, science was mainly devoted to combating 
superstition, the principal aim was rigorously to realize the dreams of 
Epicurus and Lucretius-to overthrow the idea of the gods controlling the 
Universe, to emancipate Nature from the grip of haughty lords and dark, 
mysterious forces, to demystify the Universe and face truth head on. In 
other words, knowledge for the sake of knowledge, the elevation of mankind 
with an understanding of his surrounding without need of resorting to white 
lies. 

We live in different times. Nowadays oppression does not come from powers 
making claims on behalf of divinity. The value that motivates today's world 
is called Money rather than God. The conspiracy between capitalism and 
democracy is all-consuming, their enemies have two destinies: either be 
absorbed or be eliminated. The applied sciences have always been allied 
to capitalism; they drive technology and flood the market with products 
labelled with a price. The pure sciences, or those with non-industrial 
applications, such as astronomical research, for the most part, were 
revised in terms of their driving principles; they were adapted and absorbed 
to the needs of our times. Present-day utilitarianism revolts at the idea of 
knowledge for its own sake. Even Buddhism, with its initially antimaterialist 
ideas, has been rendered into a marketable product in the book shops or in the 
form of courses on transcendental meditation. Culture has also been turned 
into a "cultural industry", to use Adorno and Horkheimer's expression. 
Scientific knowledge has become a milk cow on which to grow
fat\footnote{{\it ``Science. Heavenly goddess for some people, and an 
industrious cow which produces butter for other \\ people.''} (Goethe)}, an 
industry providing jobs to some State employees in order to make possible 
they can live with their spouses and their children in the welfare state.

Genius science is substituted for science of the masses and for a democratic 
science that advances with the rhythm of mediocrity. The stomach is put before 
the brain. Everything is bureaucratized, everything requires paperwork and to 
conform to mediocrity in order to effect a project.

\begin{quotation}
{\it ``Democratic resentment denies that there can be anything that can't 
be seen by everybody; in the democratic academy truth is subject to public 
verification; truth is what any fool can see. This is what is meant by the 
so-called scientific method: so-called science is the attempt to democratize 
knowledge - the attempt to substitute method for insight, mediocrity for 
genius, by getting a standard operating procedure. The great equalizers 
dispensed by the scientific method are the tools, those analytical tools. 
The miracle of genius is replaced by the standardization mechanism.''} 
(Norman O. Brown, ``Love's body'')
\end{quotation}

Those research ideals have been left behind. Intellectual restlessness, 
the search for truth created those colossi of knowledge who moved among 
the different fields like salmon among rapids. Today, such pirouettes have 
become impossible because knowledge has become heavy and sluggish. You will 
see an elephant sliding before you see a scientist knowing so many fields 
as our scientific forefathers did. Nowadays, a scientist has to specialize. 
Scientists have been specializing for quite a long time, but it is now a 
question of microspecialization. There are experts on cool stars, the 
Galactic bar, certain types of chemical reactions, etc. The most a scientist 
can hope to achieve is mastery of a few microspecializations, in which to 
invest their efforts or creative interests. It is hard to imagine someone 
getting into a specialization because it is his only interest, unless the 
system has sent him crazy enough to believe that his topic is the centre 
of the world. This clearly is not so. Rather, it is more a case of 
converting the scientific process into an industrialized mass-production 
system. Everybody attends to his own cog so that the system runs smoothly.

It is a treason to our scientific forefathers' ideals. Descartes gave 
science a sense for mankind as a search of truth in his ``Rules for the 
direction of the mind'', and expressed in the first rule: 

\begin{quotation}
{\it ``Thus, if somebody wants seriously to research the truth of the things, 
must not choose a peculiar science, since all of them are related among 
themselves; rather, he must only think about increasing the natural light 
of reason, not in order to solve this or that school difficulty but to get 
an understanding about life that shows us the behaviour we have to choose.''}
\end{quotation}

And, what does the scientific industry produce? The answer depends on the 
observer. From inside, we see tons of printed paper which is not read by 
anybody except some few specialists, each one about his topic. From outside 
we get a hermetic impression, such as we said ``what amazing things these 
people must be discovering! It must be so difficult and advanced that it 
not accessible to my level of understanding''. That is the impression which 
is produced among those who pay the taxes so the State will continue funding 
further research. Those who are dedicated to applied sciences have an easiest 
task because they promise technological advances. The pure sciences without 
immediate practical applications, in order not to loose the thread of State 
subventions, must also promise technological progress for the country in the 
long-term. If it is necessary, they lie. If the technological argument does 
not work, they attempt to impress people with the knowledge content. If it 
is necessary, they exaggerate. They say that a satellite or a telescope is 
going to create a revolution in astronomy, that we are going to observe the 
whole Universe and some parts of other ones,... and then the artefact arrives 
and... the revolution has been rather small. Perhaps they scrape something 
else about some galaxy which was not in our collection.

We must not deceive ourselves. The more the history advances, the more 
difficult the achievement of a relevant truth is. Newton's scientific 
activities during one year of his life, with a mere notebook and a pen, 
were more fruitful than the activities of thousands of the best actual 
scientists in their whole lifetime and with millions of euros... It seems 
that there are many writings, many data,... but in the final analysis our 
comprehension of Nature in global terms advances at a very slow and nearly 
imperceptible pace. Great efforts bear less fruit.

\begin{quotation}
{\it ``Death of science consists of the existence of nobody able to live it. But 
200 years of scientific orgies get fed up in the end. It is not the individual 
but the spirit of a culture who gets fed up. And this is manifest by sending 
to the historical world of nowadays\footnote{Decade of 1920s, when there were 
still some important discoveries in physics. Nonetheless, I think Spengler 
is right in his prediction about the decline of the scientific world, as 
well as the decline of culture in general, although it was not as soon as 
he predicted. In my opinion, he was beyond his time, and he saw the 
problems of the future in our civilization in a prophetic way.} 
researchers who are more and more small, mean, narrow and infecund.''} 
(Oswald Spengler, ``The Decline of the West'')
\end{quotation}

The fight for the economic power and social status promote fights among 
specialist from different fields rather than searching ``the truth'' 
altogether. Astronomers ask money because they are disembowelling the cosmos 
secrets; the particle physicists are disembowelling the matter secrets; the 
biologist the life secrets;... What impatient individuals who want to reveal 
all of Nature's mysteries and do not want to leave anything for the next 
generations! Some data has not still been exploited completely and we think 
about getting the next data. Fast!, before any other makes the discovery! 
Impatience has never been typical of wise people. I know well your little 
secret: will to power. In regard to this topic, Nietzsche has made a deep 
psychological analysis of men's intentions:

\begin{quotation}
{\it ``Why do we try to demonstrate the truth? Because of a larger feeling of 
power, because of the utility, because it is indispensable. Summing up, in 
order to get some advantages. But this is a bias, a signal which points that 
deep down we do not worry about the truth.''} (Nietzsche, ``The Will to Power'')
\end{quotation}

The fight among specialists from different branches is similar to that for 
defending the lands in the medieval age. The ``authorities of the matter'', 
as they call themselves, are like lords of some lands who guard fervently 
their kingdom. When an intruder tries to insert his nose in a specialty 
which is not his, he will soon receive a cohort of ``authorities'' reading 
his rights. Generally, the lands are also fenced with a language and symbols 
to be crossed only by experts. In some occasions, I would say that formalisms 
are made to frighten other people, in order to make the entrance difficult.

Saying that doing research is collaborating for the peace and fecundity of 
mankind's progress\footnote{ This was said by the king of Spain, Juan Carlos
I.} is slightly naive. Nations do not invest on research today because of 
beautiful phrases like the above one. Nations, like persons, look for prestige.
A country sends its sportsmen to the Olympic Games to win prestige, in order 
to get people to say: `sportsmen with certain nationality won a medal...', and 
then the national hymn will be played and all that. Next day, the newspapers 
publish in their pages ``our sportsmen won some medals in ...'', this ``our'' 
makes the reader feel proud to belong to his country and then he will like 
to produce for his society. In the same way, the State pays scientists, even 
non-technological ones. If they are not useful for industrial production, 
they are at least useful to produce prestige. It is very beautiful to find 
in the news: ``a scientist of our research centres discover...'', it makes 
the citizens believe he lives in a true country. There are meetings about 
science even in undeveloped countries, do they also want to collaborate for 
peace and the fecundity of mankind's progress while their citizens live 
on poverty?

The great Spanish philosopher Miguel de Unamuno, expressed in the presence of 
these events: ``let them invent!''. This expression has contained more than 
a simple rebuff. If we are interested in knowing the truth, the way is not 
through microspecialization. Let the nations invest their efforts, their 
own pride will announce the news to the world and the ideas you were 
interested will arrive at your ears. Of course, this position does not 
include neither a job nor a medal, only wisdom and prudence.

\begin{quotation}
{\it ``Do they invent things? Invent them! Electric light is here as good as in 
the place where it was invented. (...) On one hand, science with its 
applications is useful to make life easier. On the other hand, it is useful 
to open a new door for the wisdom. And are not there other doors? Have not 
we another one?''} (Unamuno, letter to Ortega y Gasset)
\end{quotation}

\begin{quotation}
{\it ``Yes, yes, I see it; a huge social activity, a powerful civilization, a 
lot of science, a lot of art, a lot of industry, a lot of morality, and then, 
when we have filled the world with industrial wonders, with large factories, 
with paths, with museums, with libraries, we will fall down exhausted near all 
this, and it will be, to whom?, was man made for science or science made for 
man?''} (Unamuno, ``Tragic Sense of Life in Men and in Peoples'')
\end{quotation}

That is, to whom? Perhaps, the historical moment when we must raise again the 
question has already arrived. Where are we going? The scientific method is 
awfully eroded. That thing with a reason for being at the beginning of the 
modern age as a promoter of positive knowledge; that later century of 
enlightenment;... all that is part of the past. Today, science is as crushed 
as the contemporary art. In words by Feyerabend in his ``Against method'':

\begin{quotation}
{\it ``Science failed to be a variable human tool to explore and change the 
world and rendered itself into a solid block of knowledge, impermeable to 
human dreams, wishes and hopes.''}
\end{quotation}

Science looses its first attractiveness, simple technical operations remain. 
Which is the thing in whose name we do research? In the name of truth? 
Of economy? Of prestige? Max Weber thinks that the dreams of a science as a 
way to the truth, or happiness, or knowledge of God, etc. are shipwrecked. 
Neither the scientist is a prophet-says Weber. Science as an amusement still 
remains but the growing pedantry and smugness limits it.

Doing research is fighting, what else human beings could do? To fight against 
the power of others or to attain our own power, that depends on us. Science 
can be a revolution or deadlocked idleness. Still waters, without hitting the 
stones along their history, tend to form bogs.

\end{document}